\documentclass[epj]{mysvjour}
\usepackage{graphics}
\usepackage{graphicx}
\begin{document}
\title{Optimization of neutrino beams for underground sites in Europe}

\subtitle{}
\author{A. Longhin\inst{1}
}                     
\institute{INFN Laboratori Nazionali di Frascati, via E.Fermi 40, 00044 Frascati, Italy;\\
on leave from CEA, Irfu, SPP, Centre de Saclay, F-91191 Gif-sur-Yvette, France.}
\date{}
\abstract{
We present an optimization procedure for neutrino beams which could be produced at CERN and aimed 
to a set of seven possible underground sites in Europe with distances ranging from 130~km to 2300~km.
Studies on the feasibility of a next generation very massive neutrino observatory
have been performed for these sites in the context of the first phase of the LAGUNA design study.
We consider specific scenarios for the proton driver (a high power proton driver at 4.5 GeV 
lor the shortest baseline and a 50 GeV machine for longer baselines) 
and the far detector (a Water Cherenkov for the shortest baseline and a LAr TPC
for longer baselines). The flux simulation profits of a full GEANT4 simulation. 
The optimization has been performed before the recent results on $\nu_e$ appearance by reactor 
and accelerator experiments and hence it is based on the maximization of the sensitivity on 
$\sin^22\theta_{13}$. Nevertheless the optimized fluxes have been widely used since their
publication on the internet (2010). This work is therefore mainly intended as a documentation 
of the adopted method and at the same time as an intermediate step towards future studies which will put the emphasis on the 
performances of beams for the study of $\delta_{CP}$.
} 
\maketitle
\section{Introduction}
\label{intro}

The feasibility of a European next-generation very massive neutrino
observatory in seven potential candidate sites located at distances
from CERN ranging from 130~km to 2300~km, has been explored within
the LAGUNA\footnote{``Large Apparatus studying Grand
Unification and Neutrino Astrophysics'', FP7 EU program} design study \cite{LAGUNA}. 
In order of increasing distance from Geneva the considered sites have been Fr\'ejus (France) 
at 130~km, Canfranc (Spain)  at 630~km, Caso (Italy) at 665~km, Sierozsowice (Poland) at 950~km,
Boulby (United Kingdom) at 1050~km, Sl\u{a}nic (Romania) at 1570~km and  Pyh\"asalmi (Finland) at 2300~km.

When coupled to very intense neutrino beams from CERN, large detectors hosted in such 
an underground site, could measure with high precision 
the mixing angle $\theta_{13}$, and eventually
determine the neutrino mass hierarchy and the existence 
of CP violation in the leptonic sector.

\begin{figure}
\resizebox{0.5\textwidth}{!}{%
\includegraphics{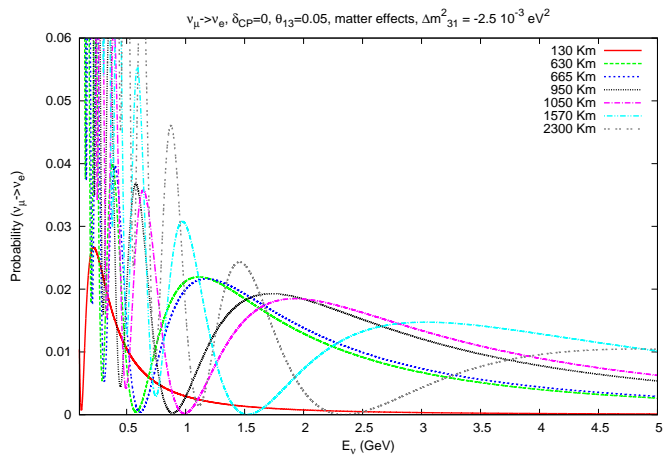}
}
\caption{The neutrino oscillation probability ${\mathcal{P}}(\nu_\mu\to\nu_e)$ for the LAGUNA baselines. 
We assume the inverted hierarchy with $\sin^22\theta_{13}$=0.05, $\delta_{CP}=0$
and the presence of matter effects.}
\label{fig:proba}
\end{figure}

The oscillation probability of the $\nu_\mu \to \nu_e$ channel is shown
as a function of the neutrino energy in Fig. \ref{fig:proba} for the considered baselines. 
The energy of the first oscillation maximum spans a wide range of energies 
for the considered baselines ranging from 0.26~MeV at 130~km up to 4.65~GeV at 2300~km,
the full sequence being $\{$0.26, 1.27, 1.34, 1.92, 2.12, 3.18, 4.65$\}$~GeV.

This parameter is crucial for the optimization of the energy spectrum of the neutrino beam
as it will be shown later. 
Neutrino spectra should cover the region where the oscillation effect is 
maximal with high statistics and low intrinsic contamination of $\nu_e$. 
The study of CP-violation 
requires to measure the oscillation probability as a function of the neutrino
energy, or alternatively to compare large samples of $\nu_e$ and
$\bar{\nu}_e$ CC events, and suffers in general from neutrino
oscillation parameters degeneracies. The possibility
to have a broad beam covering the second oscillation maximum at lower energy is
beneficial since it provides additional input useful to constraint the 
effects of mass hierarchy and the $\delta_{CP}$ phase \cite{Hasegawa}
and limits the impact of systematic errors on flux normalization
by providing spectral information.

In this work we
investigate two options for the proton driver: a high power superconducting 
proton linac at 4.5~GeV and a high power synchrotron at 50~GeV.
Concerning the detector technology we consider a 440 kt Water Cherenkov 
for the 130~km baseline and the low energy proton driver and a 100 kt LAr Time Projection Chamber (LArTPC)
at longer baselines with the high energy accelerator. Realistic designs have been proposed 
for these two detectors: the MEMPHYS \cite{MEMPHYS} and the GLACIER \cite{GLACIER} concepts.
Previous studies on a high-energy super-beam \cite{RubbiaMarzo} and a low energy super-beam \cite{Mezz03}, \cite{AJAC}, \cite{Mezz06} \cite{ICHEP2010_io} are available.
The simulation of fluxes is based on the GEANT4 \cite{GEANT4} libraries and the optimization is performed separately for 
each one of the considered baselines. The guiding line of the optimization is the final achievable sensitivity 
on $\sin^22\theta_{13}$. Furthermore a direct comparison of a high-energy and low-energy super-beam based on different accelerator scenarios has been done using of a coherent set of simulation tools.

\section{Proton drivers and detectors}
\label{sites}

Neutrino rates in conventional beams are at first approximation proportional 
to the incident primary proton beam power, hence intense neutrino beams 
can be obtained by trading proton beam intensity with proton energy.  
So two basic approaches may be considered: a relatively low proton energy 
accompanied by high proton intensity or a higher proton 
energy with lower beam current.  

In low energy neutrino beams the bulk of $\nu_e$ contamination comes from the $\pi\to\mu$ decay chain 
and only marginally from kaon decays ($\sim 10\%$ of the $\nu_e$ at 4.5 GeV proton energy). This source of background can therefore be more easily constrained due to the correlation with the dominant $\nu_\mu$ flux component from direct pion decays. In addition, the horn and decay tunnel can be kept at relatively small scales. 
Finally in the sub-GeV region most of neutrino interactions are quasi-elastic. This final state allows an easy configuration for the calculation of the parent neutrino energy and can be cleanly reconstructed also in a water Cherenkov detector. The $\pi^0$ rejection from neutral current events also benefits from the low energy regime since photons are less collinear and energetic allowing the Cherenkov ring patterns generated by $e^+e^-$ to be more resolvable. 
In order to fulfill the condition of being on the first maximum of oscillation the baseline has to be conformingly small and this offers the advantage of having a small suppression of the flux ($\sim L^{-2}$). Furthermore the determination of CP violation at small baselines is cleaner since there is almost no interplay with CP violating effects related to matter effects.

On the other hand high energy super beams associated to large baselines offer the possibility to study the neutrino mass hierarchy via the study of matter effects in the earth. The neutrino cross section scales about linearly with the energy allowing comparatively larger interaction rates at fixed flux. At high energy neutrino cross sections are free from the large theoretical uncertainties present in the low energy regime (nuclear effects, Fermi motion) which make the use of near detector compulsory for low energy super-beams. These effects also spoil the neutrino energy resolution. Furthermore the pion focusing is more efficient at high energy of the incident protons due to the more favourable Lorentz boost.
We note finally that the chance to measure both first and second maxima increases with the baseline
since, in general, the second maximum tends to fall at low energy where resolution and efficiency degrade. 

In the following we describe the assumptions for the proton drivers:
\begin{itemize}
\item A conceptual design report (CDR2) exists for the high power super conducting linac
(HP-SPL) \cite{SPLCDR}. It is foreseen as a 4 MW machine working at 5 GeV proton kinetic energy. 
At a first stage it would feed protons to a fixed target experiment to produce an intense
low energy ($\sim$ 400 MeV) super-beam. On a long time scale this machine could also be used to provide protons for the muon production in the context of a Neutrino Factory. 

\item The scenario for a high-power 50 GeV synchrotron (HP-PS2) has been 
initially proposed and discussed in \cite{RubbiaMarzo}.
A factor four in intensity is assumed compared to the baseline parameters 
defined by the PS2 working group \cite{PS2ref} (1.2 $\cdot$ 10$^{14}$ protons with a cycle of
2.4~s). A value of 3 $\cdot$ 10$^{21}$ protons on target (p.o.t.) 
per year could be achieved by doubling both the proton intensity and the
repetition rate.
\end{itemize}

\begin{table}[hbpt]
\centering
\begin{tabular}{|c|c|c|}
\hline
Parameter & HP-SPL & HP-PS2 \\
\hline\hline
$p$ kin. energy (GeV) & 5 & 50 \\
\hline
repetition frequency (Hz) & 50 & 0.83\\
\hline
$p$ per pulse (10$^{14}$) & 1.12 & 2.5\\
\hline
average power in 10$^7$s (MW) & 4 & 2.4\\
\hline
p.o.t/year (10$^{21}$) & 56 & 3\\
\hline
\end{tabular}
\caption{Parameters of considered proton drivers.}
\label{tab:pdrivers}
\end{table}

A summary of the assumed accelerator parameters is given in Tab. \ref{tab:pdrivers}. 
For the far detector we concentrated on two designs:
\begin{itemize}
\item
 The MEMPHYS water Cherenkov detector is envisaged as  
  consisting of 3 separate tanks of 65 m in diameter and 65 m height
  each (440 kt). Such dimensions meet the requirements of light attenuation
  length in (pure) water and hydrostatic pressure on the bottom
  PMTs. A detector coverage of 30\% can be obtained with about 81.000
  PMT of 30 cm diameter per tank. 
  Based on the  extensive experience of Super-Kamiokande, this technology is best
  suited for single Cherenkov ring events typically occurring at
  energies below 1 GeV.

\item
GLACIER is a scalable concept for single volume very large LAr TPC with a mass of 100~kt. The powerful imaging will 
allow to reconstruct with high efficiency electron events in the GeV range and 
above, while considerably suppressing the neutral current background mostly 
consisting of misidentified $\pi^0$s.
\end{itemize}

\section{Optimization of the focusing system}
\label{hornoptim}
The optimization of the neutrino fluxes for the CERN-Fr\'ejus baseline
with a Cherenkov detector and a 4.5 GeV proton driver has been studied 
extensively in \cite{articleSPL} so in the following we will take the optimized 
fluxes obtained in that work and focus on the optimization
of the focusing system for longer baselines assuming a LAr far detector and
a 50 GeV proton driver.

The focusing system is based on a pair of parabolic horns
which we will denote as horn (upstream) and reflector (downstream) 
according to the current terminology. This schema is the same which is
being used for the NuMI beam. 
The target is modelled as a 1 m long cylinder of graphite
($\rho=1.85$ g/cm$^3$) and a radius of 2 mm. Primary interaction in
the target were simulated with GEANT4 QGSP hadronic package.

\begin{figure}[hbpt]
\centering
\resizebox{\linewidth}{!}{%
\centering
\includegraphics{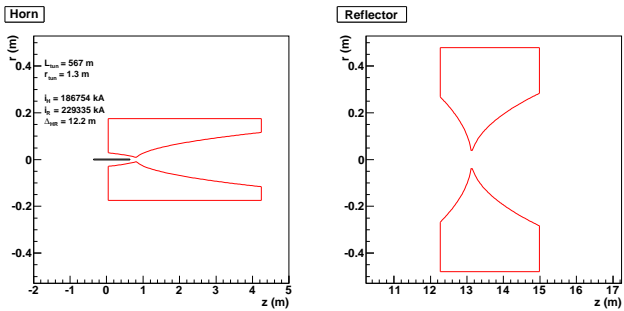}
}
\caption{Layout of the horn/reflector.}
\label{fig:hornGOOD}
\end{figure}

The optimization relies on a parametric model of the horn and reflector 
profiles. The horn radius as
a function of the coordinate along the proton beam, $r(z)$,  
has been parametrized as shown in the first row of Tab.\ref{tab:cval} 
in the three $z$ domains $[0,z_1]$, $[z_1,z_2]$, $[z_2,z_3]$.
The model contains eleven shape parameters for each magnetic lens 
($a$, $b$, $c$, $d$, $a^\prime$, $b^\prime$, $c^\prime$, $r$, $z_1$, $z_2$, $z_3$) 
which reduce to nine after requiring continuity at the points $z_1$ and $z_2$
($a$, $b$, $c$, $d$, $c^\prime$, $z_1$, $z_2$, $z_3$).
The layout of a typical configuration is shown in Fig. \ref{fig:hornGOOD}.

In addition to the parameters related to the shape of the horn and the reflector, 
additional degrees of freedom are: the distance between the horn and reflector 
($\Delta_{HR}$), the length and radius of the decay tunnel ($L_{tun}$, $r_{tun}$),
the longitudinal position of the target ($z_{tar}$) and
the currents circulating in the horn and the reflector ($i_H$, $ i_R$).

Following the approach already used in \cite{articleSPL} for the optimization of the SPL-Fr\'ejus Super Beam, we introduce, as a figure of merit of the focusing, a quantity $\lambda$
defined as the $\delta_{CP}$-averaged 99 \% C.L. sensitivity limit on
$\sin^2(2\theta_{13})$ (:=$\lambda_{99}(\delta_{CP})$) in $10^{-3}$ units
\begin{equation}
  \label{eq:lambdadef}
  \lambda = \frac{10^3}{2\pi} \int_{0}^{2\pi} 
     \lambda_{99}(\delta_{CP})\, d\delta_{CP}
\end{equation}
In the following we will denote the quantity $\lambda$ evaluated
for a specific baseline $L$ as $\lambda_L$. A sample of $10^5$ secondary meson tracks per configuration was used. Fluxes were calculated with 20 energy bins from 0 to 10 GeV. The statistical fluctuations introduced by the size of the sample have been estimated by 
repeating the simulation for the same configuration several times 
with independent initialization of the GEANT4 random number engine. 
The spread is enhanced by the presence of single events 
which can be assigned large weights. The spread on the parameters $\lambda_L$ is of the order of 3-4\%.
The sensitivity limit was calculated with the GLoBES software\cite{GLOBES} fixing a null value for $\theta_{13}$ and fitting 
the simulated data with finite values of $\sin^22\theta_{13}$ and $\delta_{CP}$ sampled in a grid of 10 $\times$ 200 points in the $(\delta_{CP},\sin^22\theta_{13})$ plane for $\delta_{CP}\in[0,2\pi]$ and $\sin^22\theta_{13}\in [10^{-2},10^{-4}]$. 
The 99\% C.L. limit was set at the values corresponding to a $\Delta\chi^2$ of 9.21 (2  d.o.f.). The normal hierarchy was  assumed in the calculation.
We assumes running periods of 2 years in $\nu$ mode 
and 8 years in $\bar{\nu}$ mode. The detector response is described in GLoBES
by assigning values for the energy resolution, efficiency and defining
the considered channels.
The parametrization of the MEMPHYS detector is the same
which was used in \cite{Mezz06}. The event selection and particle
identification are the Super-Kamiokande algorithms results.
Migration matrices for the neutrino energy reconstruction are used to
properly handle Fermi motion smearing and the non-QE event
contamination.  The reconstructed energy is divided into 100 MeV bins
while the true neutrino energy in 40 MeV bins from 0 to 1.6 GeV.  Four migration
matrices for $\nu_e$, $\nu_\mu$, $\bar{\nu}_e$ and $\bar{\nu}_\mu$ are
applied to signal events as well as backgrounds.
The considered backgrounds are $\nu_\mu^{CC}$ interactions misidentified
as $\nu_e^{CC}$, neutral current events and $\nu_e+\bar{\nu}_e$ intrinsic
components of the beam.

In the simulation of the GLACIER detector the considered backgrounds are the intrinsic $\nu_e$ and $\bar{\nu}_e$ components in the beam.  Reconstructed neutrino energy was divided in 100 MeV bins
from 0 to 10 GeV. A constant energy resolution of 1 \% is assumed for the signal and the background.

We followed two strategies in the optimization procedure
which we will describe in the following subsections.
\begin{table}[hbpt]
\centering
\begin{tabular}{|c|c|c|c|c|c|}
\hline
     $r(z)$ & 
     \multicolumn{2}{|c|}{$\sqrt{\frac{a-z}{b}}-c$} 
     & \multicolumn{1}{|c|}{$d$} 
     & \multicolumn{2}{|c|}{$\sqrt{\frac{z-a^\prime}{b^\prime}}-c^\prime$} \\
\hline
      $z$ range& 
     \multicolumn{2}{|c|}{$[0,z_1]$} 
     & \multicolumn{1}{|c|}{$[z_1,z_2]$} 
     & \multicolumn{2}{|c|}{$[z_2,z_3]$} \\
\hline
\hline
Par. & horn & refl. & Par. & horn & refl.\\
\hline\hline
$a$&85.7&100&$d$&0.9&3.9\\ 
\hline
$b$&7.0&0.135&$r$&15&40\\
\hline
$c$&0.2&0.3&$z_1$&80&97.6\\
\hline
$a^\prime$&82.2&100.&$z_2$&83.0&104.8\\ 
\hline
$b^\prime$&2.18&0.272&$z_3$&300&300\\ 
\hline
$c^\prime$&0.2&0.3&&&\\
\hline
\end{tabular}
\caption{Analytic parametrization of the horn/reflector radial profile $r(z)$ and 
central values of the parameters expressed in cm. $r$ is the conductor outer radius.}
\label{tab:cval}
\end{table}

\subsection{Fixed horn search}
In a first step we decided to fix the horn and reflector shapes (central values of Tab. \ref{tab:cval}), the tunnel geometry 
($L_{tun}=300$ m,  $r_{tun}=1.5$ m) and the circulating currents (200 kA). We then varied the relative positions of the horn, the reflector and the target.
We define the distance between the center of the target and the most upstream point of the horn as $z_{tar}$ while we indicate with $\Delta_{HR}$ the horn-reflector distance.
 After having chosen the best point in this space we did a similar exercise in the decay tunnel parameter space ($L_{tun}$,  $r_{tun}$). At first order these two couples of parameters are expected to be weakly correlated so that doing the optimization in one pair of variables after fixing a specific choice for the other pair should not have a big impact on the final result.

\begin{figure}
\resizebox{0.5\textwidth}{!}{%
 \includegraphics{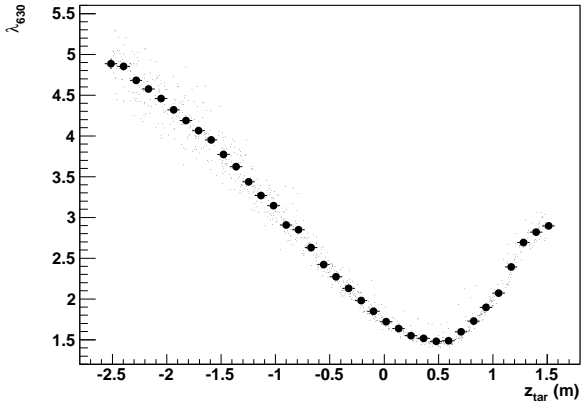}
}
\caption{Dependence of $\lambda_{630}$ on $z_{tar}$ for the fixed horn search.}
\label{fig:fhs1}
\end{figure}

The variables ($\Delta_{HR}$, $z_{tar}$) were sampled uniformly in the intervals $[0,300]$ m and $[-1.5,2.5]$ m respectively. Optimal values were then chosen for each baseline.
In general a marked dependence of $\lambda$ on the longitudinal position of the target ($z_{tar}$) is observed while variations of $\Delta_{HR}$ have a reduced impact. 
In Fig \ref{fig:fhs1} we show, taking the baseline of 630~km as an example, the dependence of $\lambda_{630}$ on $z_{tar}$ after marginalizing on $\Delta_{HR}$.
For the 630~km baseline the optimal $z_{tar}$ lies around +0.5 m, 
while for $\Delta_{HR}$ a value of 50 m was chosen. At this stage of the 
optimization the best values for $\lambda_{630}$ cluster between 1.4-1.5. 
The first two columns of Tab. \ref{tab:fixDZ} give the $\Delta_{HR}$ and $z_{tar}$ pairs 
providing the best limit for each baseline
($\lambda_{min}$, 3$^{rd}$ column).
\begin{table*}[hbpt]
\begin{center}
\begin{tabular}{|c|c|c|c||c|c|c||}
\hline
$L$ (km)& $z_{tar}^{opt}$ (m)&$\Delta_{HR}^{opt}$ (m)& $\lambda_{min}$ & $L_{tun}^{opt}$ (m) & $r_{tun}^{opt}$ (m) & $\lambda_{min}^{\prime}$\\
\hline
630&0.5&50&1.4-1.5&75&2&1.3\\
\hline
665&0.45&55&1.4-1.5&90&2.2&1.3\\
\hline
950&0&75&1.3&110&2&1.2\\
\hline
1050&-0.25&4&1.3&200&1&1.3\\
\hline
1570&-0.3&4&1.2-1.3&280&1&1.2\\
\hline
2300&-0.8&4&1.7&400&1.5&1.6\\
\hline
\end{tabular}
\end{center}
\label{tab:fixDZ}
\caption{Fixed horn shape search. Optimal values for $\Delta_{HR}$, $z_{tar}$, $L_{tun}$ and $r_{tun}$.}
\end{table*}

\begin{figure}[hbpt]
\centering
\resizebox{0.9\linewidth}{!}{%
\includegraphics{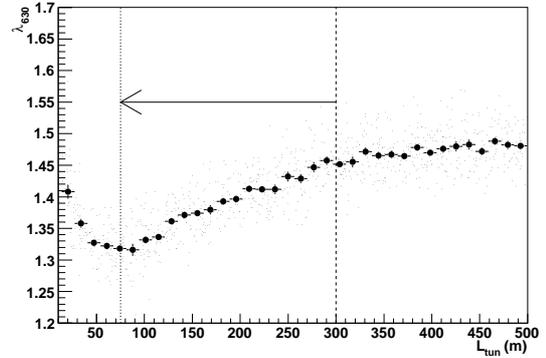}
}
\resizebox{0.9\linewidth}{!}{%
\includegraphics{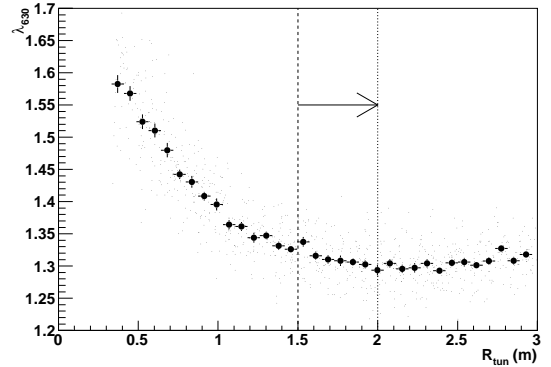}
}
\caption{Dependence of $\lambda_{630}$ on $L_{tun}$ and $r_{tun}$ in the fixed horn search.}
\label{fig:fhs2}
\end{figure}

After having fixed $\Delta_{HR}$, $z_{tar}$ to the optimal values of Tab. \ref{tab:fixDZ}, the tunnel length $L_{tun}$, previously fixed at 300 m, was sampled uniformly between $[10, 500]$ m keeping $r_{tun}$ fixed at 1.5 m. The optimized values for $L_{tun}$ are given in Tab. \ref{tab:fixDZ} (4$^{th}$ column). In the case of $L=$~630~km a gain of order 20\% is visible in Fig. \ref{fig:fhs2} (left) decreasing $L_{tun}$ from a 300 to 75~m. 

The $r_{tun}$ was then sampled in $[0,3]$ m having fixed the optimal tunnel length. The right plot of Fig. \ref{fig:fhs2} shows that 
an improvement is obtained increasing $r_{tun}$ to 2 m. The optimized values for $r_{tun}$ are shown in Tab. \ref{tab:fixDZ} (5$^{th}$ column). The values of $\lambda$ obtained after the tunnel optimization ($\lambda_{min}^\prime$) are shown in the 6$^{th}$ column of Tab. \ref{tab:fixDZ}. The variation between $\lambda_{min}$ and $\lambda_{min}^\prime$ 
shows that the tunnel optimization is particularly effective for the short baselines for which
the initial geometry was not appropriate.
\subsection{General search}

In order to understand how limiting is the choice of fixing the shape of the conductors and the
circulating currents we performed a high statistics general scan of the configurations allowing also these parameters to vary. Also the parameters which were previously optimized were varied since in general changing the shape of the conductors and the currents we do not expect the same optimization to be valid any more.

The shape parameters of the horn and the reflector were sampled with uniform distribution 
within 50 \% of  their central values given in Tab. \ref{tab:cval}. The other parameters were
sampled with uniform distribution in the ranges of Tab. \ref{tab:pvar}.
\begin{table}[ht]
\centering
\begin{tabular}{|c|c|c|c|}
\hline
Parameter & interval & Parameter & Interval\\
\hline\hline
$L_{tun}$&[200,1000] m& $r_{tar}$&2 mm\\
\hline
$r_{tun}$&[0.8,2] m& $\Delta_{HR}$&[4,300] m\\
\hline
$z_{tar}$&[-2.5, 1.5] m& $i_H$, $i_R$ &[150,300] kA\\
\hline
\end{tabular}
\caption{Focusing system parameters not related to the horn-reflector shapes.}
\label{tab:pvar}
\end{table}
The inclusive distributions of the input parameters were compared to the corresponding
ones for the sub-sample with $\lambda < $1.5 in order to pin down the variables which are 
effective in producing good results. Despite the smearing effect introduced by the simultaneous variation of many correlated variables, a visible trend is observed for $z_{tar}$ which exhibits 
a strong correlation with $\lambda$. 
It is clear that putting the target more and more upstream with respect to the horn, is mandatory to get good exclusion limits, as far as the baseline increases.

\begin{figure}[hbpt]
\centering
\resizebox{0.9\linewidth}{!}{%
\includegraphics{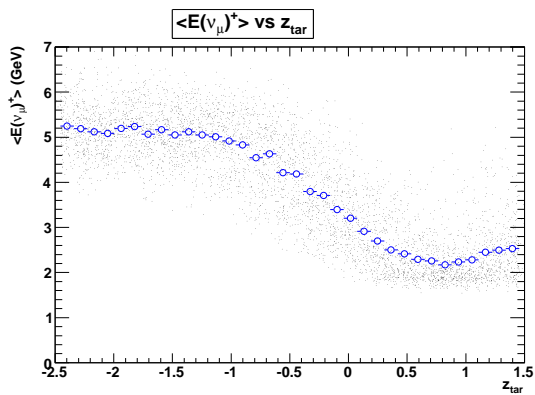}
}
\caption{Correlation between the longitudinal position of the target with respect to the horn and 
$\langle E_{\nu_\mu}\rangle$.}
\label{fig:EZ}
\end{figure}

The correlation between the longitudinal position of the target with respect to the horn and the mean energy of the $\nu_\mu$ spectrum ($\langle E_{\nu_{\mu}}\rangle$) is shown in Fig. \ref{fig:EZ}. Putting the target upstream, high energy pions, which are typically produced at small angles,  are preferentially focused resulting in a high energy neutrino spectrum.

\begin{figure}[hbpt]
\centering
\resizebox{0.9\linewidth}{!}{%
\includegraphics{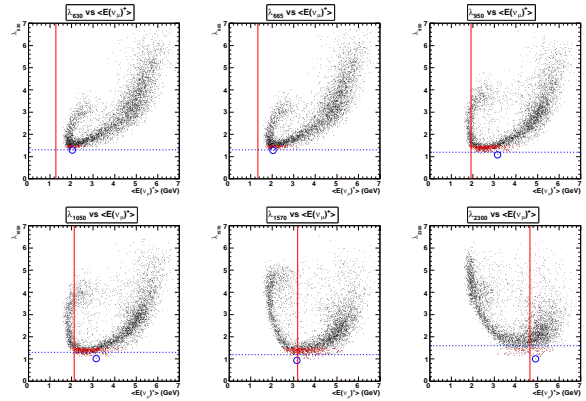}%
}
\caption{Correlation between the figure of merit $\lambda$ and $\langle E_{\nu_{\mu}}\rangle$ (positive focusing).}
\label{fig:fluxesLEHE}
\end{figure}

The correlation between $\langle E_{\nu_\mu}\rangle$ and $\lambda$ is shown
in Fig. \ref{fig:fluxesLEHE}. 
In general the optimal energies tend to roughly follow the position of the first oscillation maximum (red vertical lines in the plots). 
Mean energies below 2 GeV are difficult to get with a 50 GeV proton beam. A possible solution, which has not be considered in this work, could be to go towards an off axis beam for baselines lower than 600~km. The horizontal blue lines show the lowest values for $\lambda$ obtained with the
previous fixed horn shape search. 

The achieved performance is not drastically improved by the general search though some gain appears
for $L>1000$~km. Blue markers highlight the configuration providing the best limit for each baseline. It turns out that the same configuration provides the best limit both for 630 and 665~km and the same happens for 950-1050 and 1570~km. 
Given the limited improvement, we decided to stick with the best candidates obtained with the fixed horn shape search. This choice is also motivated by the fact that choosing the configurations with the minimum $\lambda$ has the disadvantage of being sensitive to statistical fluctuations. 

\section{Optimized fluxes}
\label{results}
\begin{figure}[hbpt]
\centering
\resizebox{\linewidth}{!}{%
\includegraphics{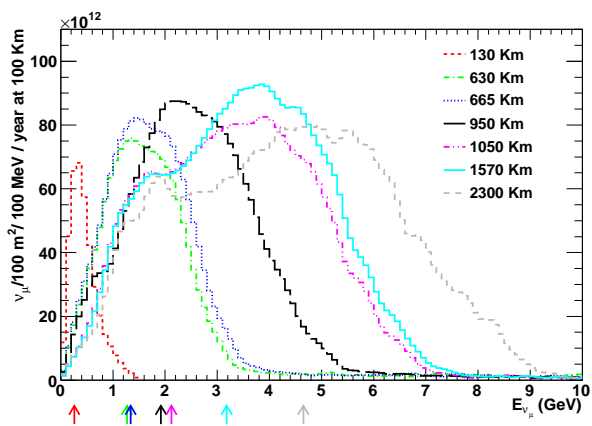}
}
\caption{Neutrino fluxes at 100~km for the systems optimized with the fixed horn shape search.
 The energies of the oscillation maximum for each baseline (Fig. \ref{fig:proba}) are indicated with vertical lines having the same color as the corresponding spectrum. The integral fluxes in units of $10^{15} \nu_\mu / 100~\rm{m}^2 / \rm{year}$ are 0.38, 1.59, 1.81, 2.69, 3.56, 3.93 and 4.48 in order of increasing baseline.}
\label{fig:fluxesOPT}
\end{figure}

The $\nu_\mu$ fluxes obtained with the optimized focusing setups according to the fixed shape search are shown in Fig. \ref{fig:fluxesOPT} at a reference distance of 100~km. Fluxes are publicly available on the internet \cite{fluxesWEB}.
The flux increase as the mean energy increases can be intuitively explained considering that high energy pions are easier to focus since they naturally tend to 
emerge from the target in the forward direction and the neutrinos they produce have a higher chance to be in the far detector solid angle also thanks to the effect 
of the Lorentz boost.
Un-oscillated interaction rates are given in Tab. \ref{tab:rates}.
It can be noted that considerable samples of $\tau$ events becomes
collectable with the fluxes optimized for the longer baselines.

\label{limits}

\begin{table*}[hbpt]
\begin{center}
\begin{tabular}{|c|c|c|c|c|c|c|c|c|}
\hline
& \multicolumn{4}{|c|}{$\nu$ run} & \multicolumn{4}{|c|}{$\bar{\nu}$ run}\\
\hline
(km) &
$\nu_\mu$ &
$\nu_e $ &
$\nu_\tau$ &
$\frac{\nu_e+\bar{\nu_e}}{\nu_\mu+\bar{\nu_\mu}}$ ({\small{\%}})&
$\bar{\nu}_\mu$&
$\bar{\nu}_e$&
$\bar{\nu}_\tau$ &
$\frac{\nu_e+\bar{\nu_e}}{\nu_\mu+\bar{\nu_\mu}}$ ({\small{\%}})\\
\hline
130 & 41316  & 174  &/& 0.42 & 5915 & 15 &/& 0.42\\
630 & 36844  & 486  &28& 1.5 & 13652 & 157 &11& 2.0\\
665 & 38815  & 516  &28& 1.5 & 14287 & 158 &11& 2.0\\
950 & 37844 & 349  &40& 1.0 & 14700 & 107 &15& 1.3\\
1050 & 51787  & 314  &148& 0.64 &21728 & 88 &65& 0.60\\
1570 & 26785  & 174  &170& 0.67 & 11184 & 47 &73& 0.57\\
2300 & 17257  & 110 &377& 0.67 & 7577 & 32 &172& 0.60\\
\hline
\end{tabular}
\end{center}
\caption{Charged current event rates with the optimized fluxes. Fractions are expressed as a percentage. Numbers are normalized to a detector mass of 100 kt and a running time of one year corresponding to $3\cdot 10^{21}$ p.o.t. for the 50 GeV proton driver
and $56 \cdot 10^{21}$ p.o.t. for the 4.5 GeV option. 
}
\label{tab:rates}
\end{table*}
\section{Conclusions}
As it was shown in \cite{ICHEP2010_io}, using the fluxes optimized
with the procedure described above, the ``discovery potential'' for $\theta_{13}$ 
turned out to be, at first order, almost independent of the baseline. 
Performances of high- and low-energy super-beams are 
comparable if we assume for both a 5\% systematic error on the fluxes. Concerning the 
high-energy super beam, better results are obtained for intermediate baselines 
from 950 to 1570~km even though the difference is not marked. 
This merit factor, despite being obsolete after the recent experimental 
results, is probably still a reasonable indicator of the precision with which
$\theta_{13}$ could be measured with these configurations.

This result could be achieved by a systematic tuning of a few basic parameters
of the focusing system: the horn-reflector distance,
the target position and decay tunnel geometry.

An exhaustive discussion of the physics potential in terms of CP violation and mass
hierarchy obtainable with the fluxes whose optimization is described in this work, has been
recently developed in \cite{Pilar}, \cite{AndreEtAl}, \cite{Pilar1}, \cite{San} and \cite{nuturn}.

By adopting a suitable re-definition of the figure of merit, the approach followed in this study 
could in the future be specialised to the need for optimal sensitivity on the CP violating effects 
under the light of the recent measurement of $\theta_{13}$.

\section{Acknowledgements}
I would like to thank M. Zito for scientific advice and reading of this
manuscript and A. Meregaglia for kindly providing the original GLoBES 
description of the LAr detector.
I acknowledge the support from the European Union under the European Commission
Framework Programme 07 Design Study EUROnu.

\end{document}